\documentclass[english]{article}
\usepackage[latin9]{inputenc}
\usepackage{geometry}
\usepackage{color}
\usepackage{float}
\usepackage{textcomp}
\usepackage{amstext}
\usepackage{graphicx}

\makeatletter

\DeclareFontEncoding{LGR}{}{}
\DeclareRobustCommand{\greektext}{%
  \fontencoding{LGR}\selectfont\def\encodingdefault{LGR}}
\DeclareRobustCommand{\textgreek}[1]{\leavevmode{\greektext #1}}
\ProvideTextCommand{\~}{LGR}[1]{\char126#1}

\DeclareTextSymbolDefault{\textquotedbl}{T1}
\providecommand{\tabularnewline}{\\}

\newcommand{\lyxaddress}[1]{
	\par {\raggedright #1
	\vspace{1.4em}
	\noindent\par}
}

\date{}

\makeatother

\usepackage{babel}
\begin{document}
\title{Unsupervised real-world knowledge extraction via disentangled variational
autoencoders for photon diagnostics }
\author{Gregor Hartmann\textsuperscript{1,{*}}, Gesa Goetzke\textsuperscript{2},
Stefan D\"usterer\textsuperscript{2}, Peter Feuer-Forson\textsuperscript{1},
\\
 Fabiano Lever\textsuperscript{3}, David Meier\textsuperscript{1,4},
Felix M\"oller\textsuperscript{1}, Luis Vera Ramirez\textsuperscript{1},
\\
Markus Guehr\textsuperscript{3}, Kai Tiedtke\textsuperscript{2},
Jens Viefhaus\textsuperscript{1} and Markus Braune\textsuperscript{2}}
\maketitle

\lyxaddress{\textsuperscript{1}Helmholtz-Zentrum Berlin f\"ur Materialien und
Energie GmbH, Albert-Einstein-Strasse 15, 12489 Berlin, Germany \\
\textsuperscript{2}Deutsches Elektronen-Synchrotron (DESY), Notkestrasse
85, 22607 Hamburg, Germany\\
\textsuperscript{3}University of Potsdam, Institut f\"ur Physik
und Astronomie, Karl-Liebknecht-Strasse 24/25, 14476 Potsdam-Golm,
Germany\\
\textsuperscript{4}Intelligent Embedded Systems, University of Kassel,
Wilhelmsh\"oher Allee 73, 34121 Kassel, Germany\\
\textsuperscript{{*}}Corresponding author: gregor.hartmann@helmholtz-berlin.de}
\begin{abstract}
We present real-world data processing on measured electron time-of-flight
data via neural networks. Specifically, the use of disentangled variational
autoencoders on data from a diagnostic instrument for online wavelength
monitoring at the free electron laser FLASH in Hamburg. Without a-priori
knowledge the network is able to find representations of single-shot
FEL spectra, which have a low signal-to-noise ratio. This reveals,
in a directly human-interpretable way, crucial information about the
photon properties. The central photon energy and the intensity as
well as very detector-specific features are identified. The network
is also capable of data cleaning, i.e. denoising, as well as the removal
of artefacts. In the reconstruction, this allows for identification
of signatures with very low intensity which are hardly recognisable
in the raw data. In this particular case, the network enhances the
quality of the diagnostic analysis at FLASH. However, this unsupervised
method also has the potential to improve the analysis of other similar
types of spectroscopy data.
\end{abstract}

\section*{Motivation }

\subsubsection*{SASE-FEL challenge}

Free electron lasers (FEL) enable atomic and molecular science in
the femtosecond to attosecond regime by creating highly intense photon
pulses on that time scale. However, FELs which are based on the principle
of self-amplified spontaneous emission (SASE)~\cite{SASE,SASE2},
such as FLASH~\cite{FLASH}, produce spatial, spectral and temporal
pulse properties that are strongly fluctuating from pulse to pulse.
Hence, a reliable photon diagnostic on a single-shot basis is essential
for sound data analysis of scientific user experiments performed at
such facilities. Post-experiment sorting of recorded data with respect
to different properties, such as intensity or wavelength, can reveal
signatures of physical processes otherwise obscured or even hidden
in the data sets. A number of diagnostic instruments at FELs are used
to measure the photoionisation of gas targets, such as the Gas Monitor
Detector (GMD)~\cite{GMD,GMD2} for measurement of absolute pulse
energy, THz-streaking~\cite{THZstreaking,THZstreaking2} for determination
of the photon pulse time structure~\cite{ANGULAR_STREAKING}, as
well as the online photoionisation spectrometer OPIS~\cite{OPIS,OPIS_PCA}
(see Fig.~1) and the so-called cookie-box~\cite{ANGULAR_STREAKING,MrCoffee}
which use photoelectron spectroscopy to get information about the
spectral distribution of the FEL radiation. These diagnostic methods
have the advantage that they can be designed to be almost completely
non-invasive. In a photoionisation process, due to the high FEL intensity,
a significant space charge~\cite{OPIS_PCA} can be created in the
ionised gas target in the interaction region of the instruments. This
space charge even accumulates for high FEL pulse repetition rates,
since the created target gas ions cannot dissipate fast enough by
Coulomb repulsion or be replenished with fresh, unionised atoms before
the next FEL pulse arrives. For instruments based on photoelectron
spectroscopy, such as OPIS, space charge can distort the diagnostic
measurement because it alters the kinetic energy distribution of the
photoelectrons. To minimise such space charge-induced detrimental
effects OPIS is operated at low target gas pressures. For this reason,
OPIS' single-shot spectra usually show low count rates and consequently
photolines comprise only a small number of single-electron events,
appearing as spikes in the spectrum, which are not clearly distinguishable
from random noise spikes (see Fig.~1). In order to obtain meaningful
wavelength results, a moving average scheme over variable time intervals
is usually applied. Hence, reliable shot-to-shot information, which
is important for experiments, could not be provided in the majority
of cases in the past. We here present a method to reveal the photon
properties in single-shot resolved mode, despite the low statistics,
by employing artificial intelligence that takes advantage of a special
type of autoencoder, which represents the data obtained by the diagnostics
device in a compressed and comprehensible way.

\subsubsection*{AI approach}

Traditional analysis methods like principal component analysis (PCA)
are robust and have proven their capability in various applications~\cite{OPIS_PCA}
but can be limited by two main issues: a) The method is linear und
thus intrinsically unable to describe non-linear effects and b) the
representations of the data (the principal components and their scaling
factors) are not necessarily easy to interpret. Scaling well with
high dimensionality and being able to describe non-linear effects,
neural networks became popular during the last decades as a powerful
analysis tool in all categories of science~\cite{NN_overview}. Autoencoder
(AE) networks~\cite{AE} built by layers of neurons are capable of
compressing data to lower dimensionality, the so-called latent space.
While a 1-layer AE network is equivalent to a PCA analysis~\cite{NN_overview},
problems of higher complexity and with non-linear effects can be handled
by adding multiple layers of neurons to the encoder and decoder. When
using such a network, the latent space representation cannot typically
be easily used for knowledge extraction and has to be further processed
in order to transform it into parameters that humans can interpret.
This can be done, for instance, with another neural network. However,
this process requires the setting of labels for training the network,
i.e. attribution of the actual values of certain physical properties
at the time of the measurement to the recorded data, which in our
case as well as in many other applications are not available. Variational
autoencoder~\cite{VAE,VAE2} networks (VAE) perform a sampling operation
on a mean and standard deviation vector in the dimensional bottleneck
of the network. By forcing these two vectors to be close to a normal
distribution by the use of an additional term in the loss function,
one creates a representation with a given value range and variation.
By varying the latent space within these limits it is possible, with
the decoder part of the network, to create artificial data samples
which represent possible measurement results. This idea was implemented
by so-called $\beta$-VAE-networks~\cite{BETA-VAE} in which the
disentanglement-term in the loss function is scaled by a factor, called
$\beta$. Thus, it is possible to balance the weight between a perfect
reconstruction (i.e. mean-square-error deviation of the raw and the
reconstructed data) and perfect disentanglement of the latent space
vector components, creating a compromise between the disentanglement
($L_{\text{dis}}$) and reconstruction quality ($L_{\text{rec}}$),
both represented in the overall loss function ($L_{\text{all}}$):

\begin{equation}
L_{\text{all}}=L_{\text{rec}}+\beta\cdot L_{\text{dis}}\label{eq:Loss}
\end{equation}

Generally, finding the best absolute value of $\beta$ is challenging~\cite{BETA-VAE,BETA_value}.
$\beta$ strongly depends on the data, i.e. on the noise level, the
size and shape of the region of interest, and on what measure is used
to evaluate the reconstruction quality.

\section*{Experiment}

\subsubsection*{FLASH and OPIS}

FLASH~\cite{FLASH} operates in a so-called burst mode pattern, generating
bunch trains with a burst repetition rate of 10~Hz. Each bunch train
consists of up to several hundred single photon pulses, depending
on the bunch repetition rate of up to 1~MHz. At FLASH2~\cite{FLASH2},
pulse energy and pulse duration range over 1-1000~\textmu J and 10-200~fs,
respectively, covering a wavelength range of 4-90~nm. For online
FEL wavelength monitoring with OPIS (see Fig.~1, for details see
\cite{OPIS}) a noble gas target, in our study neon (gas pressure
$4.4\cdot10^{-7}$~mbar), introduced into the interaction chamber
is ionised by the FLASH pulses. The kinetic energy $E_{kin}$ of the
generated photoelectrons is measured by four independently working
electron time-of-flight spectrometers (eTOF). With the knowledge of
the binding energy $E_{bin}$ of the excited orbitals, for our study
neon 2p and 2s, one can calculate the photon energy $E_{pho}$ via

\begin{equation}
E_{pho}=E_{kin}+E_{bin}\label{eq:Photoeffect}
\end{equation}

In the eTOF, the photoelectrons travel along a drift tube of 309~mm
in length and are then detected by microchannel plate (MCP) detectors.
Retarding voltages can be applied to the drift tubes in order to decelerate
the photoelectrons and therefore increase the energy resolution of
the eTOF spectrometers. 

\begin{figure}[H]
\begin{centering}
\includegraphics[scale=0.55]{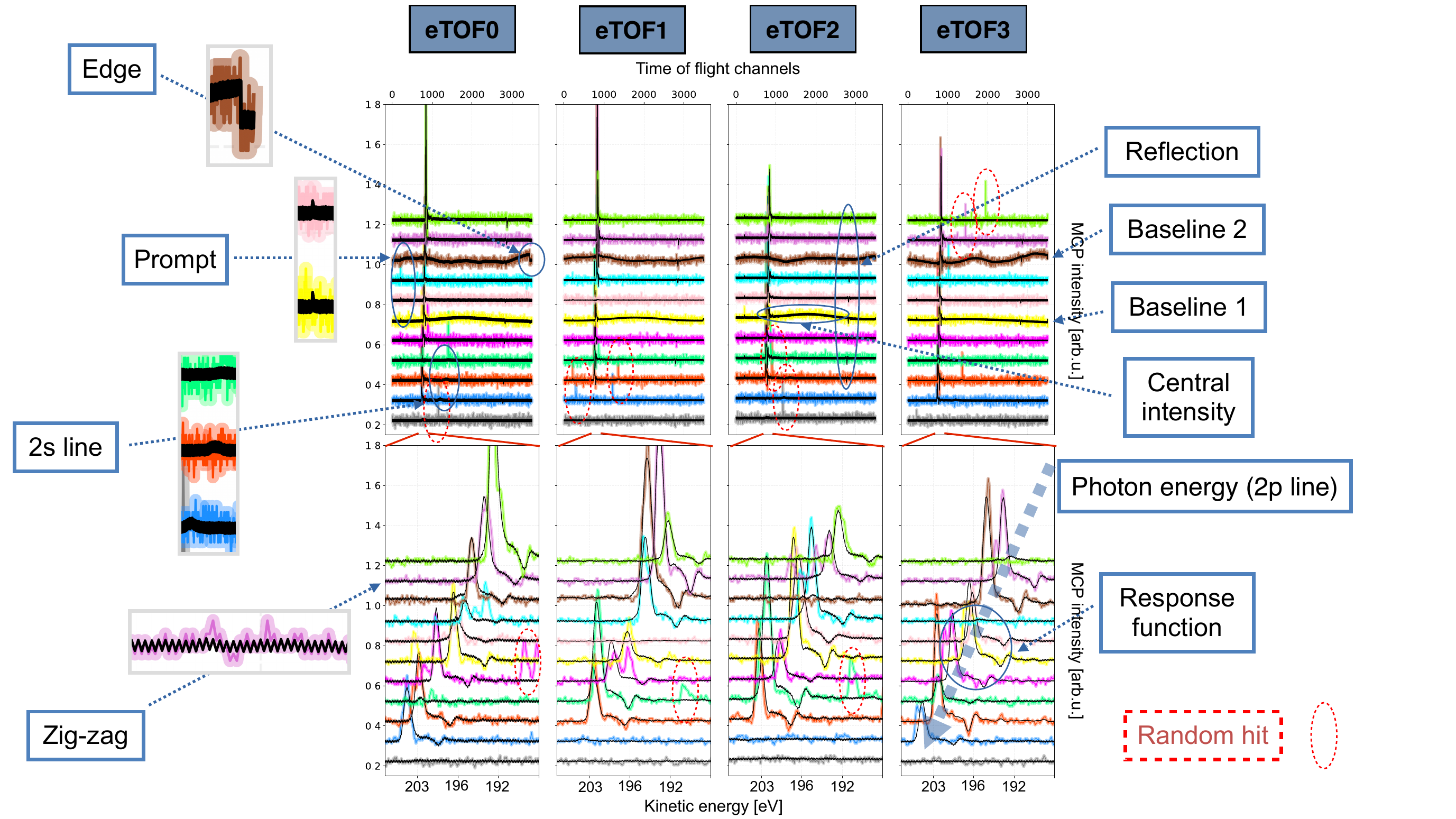}
\par\end{centering}
\caption{Eleven representative single-shot time-of-flight spectra (samples)
obtained by the four OPIS electron spectrometers (eTOF 0-3): The grey
traces at the bottom show no photo electron signal, whereas the other
ten traces above contain Ne 2p photoelectron lines with successively
longer time-of-flights, indicating decreasing FEL photon energy. The
raw data is shown by coloured bold plots while the reconstruction
of the corresponding samples is depicted by black thin lines. For
better visibility the base lines are separated by a vertical offset
of 0.1. The four upper panels show the full electron time-of-flight
spectra, which is the input of the neural network, while a zoom-in
of the corresponding region of interest, where the 2p line is expected,
is presented in the lower four panels. The zoom-in axis is converted
to kinetic energy of the photoelectron. Main features of the traces
are labelled, such as peak position, random hits, baseline disturbances,
a zig-zag structure, the prompt, 2s and 2p line, electronic reflection
due to impedance mismatch on cable connection and the corresponding
detector response function. These are reconstructed (apart from the
random hits) and encoded in the latent space. The magnified insets
represent features that are difficult to see in full scale. All scales
are linear.\label{fig:1}}
\end{figure}

\subsubsection*{Data}

Time traces of the amplified signals from the MCP detectors are recorded
by means of fast analog-to-digital converters (ADCs) with a sampling
rate of 7~GS/s and 8-bit vertical resolution. Each single-shot spectrum
consists of 3500~ADC~channels and the aggregate of the four eTOF
spectra represents one training data sample with a dimensionality
of $4\cdot3.5\text{k=14k}$ (including only an estimated number of
electrons ranging from 0-20). Some examples are presented in Fig.~1.
The intensity of the photoelectron lines in the recorded TOF spectra
are comparable in all four eTOFs, being on average within 15~\% of
the amplitude (standard deviation). However, in single-shot spectra
the photoline intensities vary significantly between the four eTOFs
due to statistical effects. Fig.~1 depicts a series of normalised
single-shot data corresponding to different values of the photon energy
of the FEL radiation, for varying time of flight of the neon 2p electrons.
A time frame of continuous wavelength monitoring is chosen in which
the OPIS operation parameters (gas target, chamber pressure, spectrometer
retardation) remained unchanged. In this time interval, the FEL photon
energy was scanned between 214~eV and 226~eV with a given irregular
pattern. In OPIS, neon was used as a target gas and the retarding
voltage was set to 170~V, resulting in a final reduced kinetic energy
of 22.4 to 34.4~eV and 0.0 to 7.5~eV of the detected 2p and 2s photoelectrons,
respectively. Roughly 40~million samples were recorded. 

\subsubsection*{Neural network}

The ultimate goal is to train a network that delivers all desired
information in a low dimensional latent space, i.e. each latent space
component should represent a property of the underlying core principle
that can be interpreted by the human mind and therefore can be directly
used as information for the experiments. For the loss function, mean-squared-error
(MSE) is used as a criterion for reconstruction quality. The disentanglement
is described by the Kullback--Leibler (KL) divergence~\cite{KL}
of the mean and standard deviation vector compared to a normal distribution.
In order to automatically stay within the value range of {[}0,1{]}
the output layer is activated with a sigmoid function. In order to
optimise the hyperparameters of the neural network about 700 different
networks were trained. The best performance was achieved with fully
connected and Mish-activated~\cite{MISH} layers with the decoder
and encoder consisting of 5 and 4 layers, respectively. Batch sizes
of 252 were used in combination with the Adam optimiser~\cite{ADAM}
and a scheduled decreasing learning rate ranging from $10^{-5}$ to
$10^{-7}$ throughout 25k epochs. The optimised value of $\beta$
is 0.034. Of the 40~million data samples recorded in total, 33~million
were used for training, 1 million for validation and the remaining
6 million represent the test data utilised outside of the training
process. The best performance of the encoder and decoder is achieved
when the layers are chosen such that the dimensionality is reduced
with the same factor for each layer, which means for 5~layers and
a 12-dimensional latent space, called $z$, the dimensionalities of
the layers are 

\begin{equation}
14000\rightarrow2824\rightarrow579\rightarrow117\rightarrow24\rightarrow12\label{eq:NN-architecture}
\end{equation}

The step from 24 to 12 is the sampling operation. The decoder is the
mirrored version of the encoder excluding the sampling operation.
The number 12 was derived by training a network starting with only
a one-dimensional $z$ and then successively increasing the size of
the dimensional bottleneck. For a size larger than 12, the final loss
value failed to significantly improve. We will use the notation $z=\left\{ z_{0},z_{1},z_{2},...,z_{11}\right\} $
to address the individual components $z_{i}$ of the latent space.

\section*{Results}

\subsubsection*{Creating labels}

The aim of OPIS measurements is to reveal values of certain physical
quantities. To analyse whether the network found a latent space representing
those quantities, labels are created by conventional analysis performed
on the raw data. To provide reliable labels the data has to meet specific
criteria, which is only applicable for a small fraction of the available
data. For example, for the flight time of the photoelectrons, i.e.
the photoline position on the TOF-scale (referred to with the label
$T_{0,1,2,3}$), a conventional least-square line profile fit analysis
of the strongest peak in each of the four eTOFs has been performed.
Here, the criteria for discrimination of a valid photoelectron line
feature from noise or random electron hits was defined such that a)
the peak amplitude has to be larger than a threshold for minimum intensity
(0.5 on the scale in Fig.~1) and b) the peak centre positions have
to lie within a small TOF range (15 TOF-channels). Applying this filter
reduces the size of the test data dramatically, but returns high quality
data. Roughly 3~\% of the data fulfills this criteria and can contribute
to the comparison between the labels and the latent space. The labels
for the individual intensities for each eTOF called $I_{0,1,2,3}$are
additionally created in the process of the peak fit procedure. The
beam position of the FEL in the plane perpendicular to the propagation
axis is also fluctuating. In order to have a robust and simple label
for these pointing variations, the 2p electron time-of-flight difference
is calculated, resulting in $P_{02}$ (eTOF0 compared with the opposite
positioned eTOF2) and $P_{13}$ (eTOF1 compared with the opposite
positioned eTOF3). This is explained in detail in the supplementary
information (SI). The ``Baseline~1'' disturbance $B_{1}$ can be
identified by evaluating eTOF0 regarding discontinuities, i.e. the
sharp \textquotedbl edge\textquotedbl{} feature at high time-of-flight
values. It is identified by calculating the sum of the intensities
of 40~ADC-channels before the edge divided by 40 after the step in
the trace baseline. The second disturbance $B_{2}$ (see ``Baseline~2''
and ``Central intensity'' in Fig.~1) is a broader feature covering
the central part of each TOF-spectrum. It is identified and labelled
by summing up the central part of the spectra which is then divided
by the mean of the data in spectral regions at the beginning and the
end of the spectrum. For the data acquisition the effective sampling
rate of 7~GSamples/s is achieved by time-interleaving four ADC chips,
sampling with 1.75~GS/s each. We realised that the network was encoding
a correlation directly pointing to systematic interleaving deficiencies:
In large parts of the data, the gain of the respective interleaved
ADCs of each eTOF channel is not identical which creates a characteristic
zig-zag-structure in the data (see zoom-in and ``zig-zag'' in Fig.~1).
This can be easily labelled by adding all odd and all even ADC channels
separately and then dividing these two sums, resulting in the labels
$L_{0,1,2,3}$. For the photon energy, one OPIS-independent value
is the set wavelength parameter $\lambda_{FEL}$ which only represents
the nominal wavelength corresponding to the FLASH accelerator and
undulator setup. The real FEL wavelength can have a certain offset,
mainly due to two factors: Firstly, the electron beam energy in the
undulator section may deviate from the energy value measured in the
accelerator section due to beam steering components such as the FLASH2
extraction and bunch compression chicanes \cite{FLASH2}. Secondly,
the electron beam orbit can deviate from the nominal orbit in the
undulator section, especially if the variable gap undulators are tuned
for wavelength scans. Furthermore, the wavelength fluctuates due to
the SASE process within a bandwidth of typically $\sim$1~\%~\cite{FLASH},
which in our case corresponds to a photon energy bandwidth of about
2~eV. Therefore, the label $\lambda_{FEL}$ is an \textquoteleft estimated\textquoteright{}
label with only moderate significance for the single-shot photon energy.
Additionally, a magnetic bottle experiment \cite{magneticBOTTLE}
was performed in parallel with our study and its data is used as a
cross reference for the wavelength, which is presented in the SI.

\subsubsection*{Reconstruction and interpretation of the latent space}

\begin{figure}[h]
\begin{centering}
\includegraphics[scale=0.45]{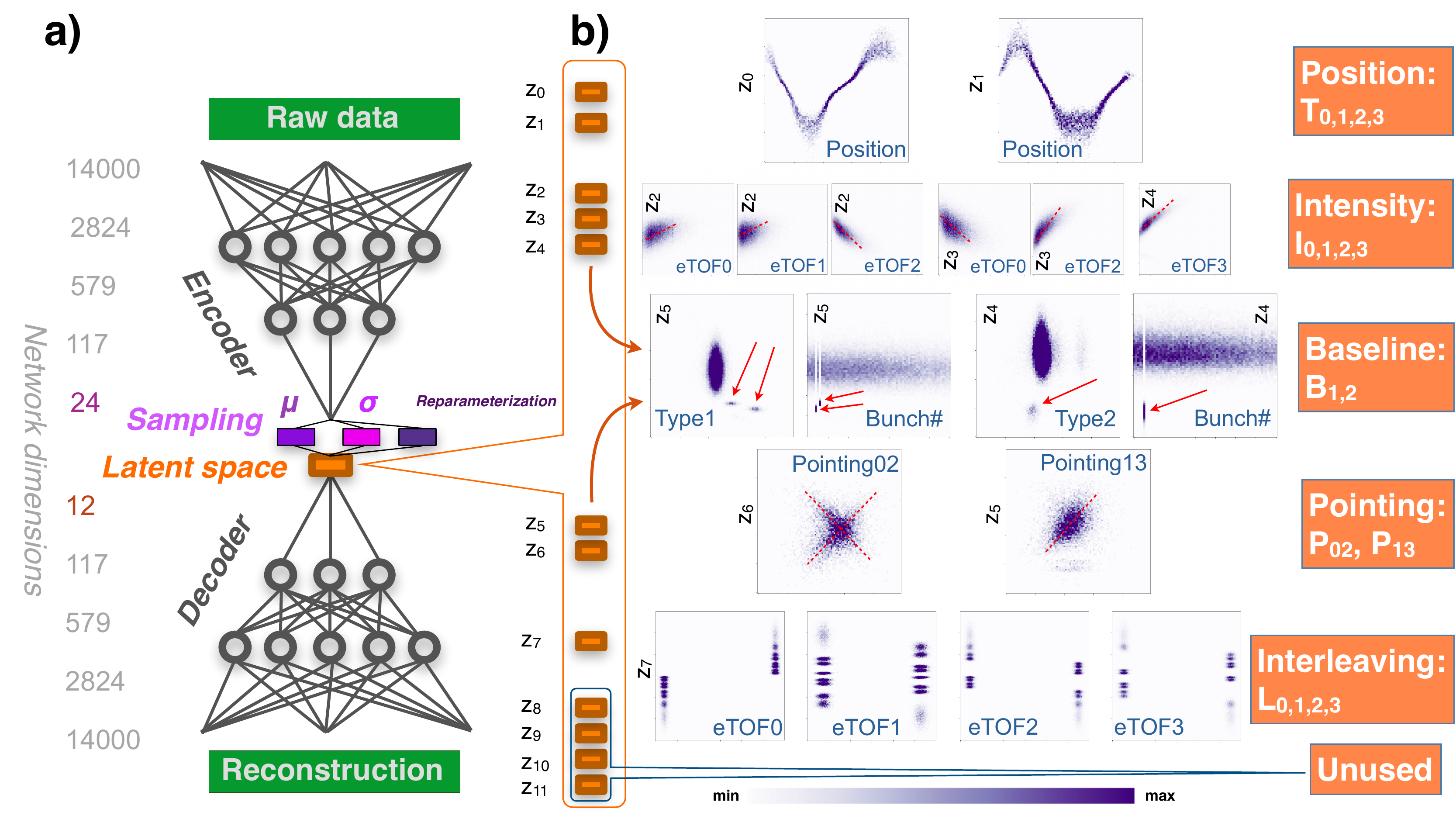}
\par\end{centering}
\caption{The structure of the $\beta$-VAE-network~(a) and the unsupervised
encoding of the underlying core principle~(b), i.e. position, intensity,
baseline, pointing and interleaving, is shown. The density plots represent
the dependency of the latent space vs. the labels, which were derived
by traditional data analysis using the high quality data (3\% of the
data set). The values in the corresponding axis ($z_{i}$ and labels)
are min-max-normalised for the processed test samples. All scales
are linear. \label{fig:2}}
\end{figure}

These labels, that result from the aforementioned feature engineering
process, are compared with the $z_{i}$ values that the network derives
for the data in Fig.~2. The reconstruction quality (black curves
in Fig.~1) is impressively high for a 12-dimensional bottleneck.
The network finds the correct position of the 2p photoelectrons, it
reconstructs the individual MCP-response function for each of the
4~eTOFs, it discards random uncorrelated events and is also able
to reproduce the baseline disturbance. In addition to these findings,
the neon 2s line is contained in the reconstruction only in cases
when the photon energy is in fact sufficiently high enough to overcome
the used retarding voltage of the flight tubes. Given that for our
data the ionisation cross section is $\sim$5 times lower for Ne~2s
compared to Ne~2p in the photon energy range of 214~eV to 226~eV
and that the 2s photoline intensity spreads over a larger TOF interval,
this is an impressive result \cite{Neon_cross-section}. Ne~2s signatures
can hardly be identified in the raw data by eye or using conventional
analysis methods. Equally impressive is the reconstruction of the
so-called prompt signal, which is created by scattered photons hitting
the MCPs and hence produces another tiny peak feature at a fixed TOF-position.
This signal marks the reference t=0 for the determination of the photoelectron
flight time and is therefore of high importance. 

A critical piece of information for most experiments at SASE-FEL sources
is the single-shot central photon energy. In OPIS measurements it
corresponds to the peak position in the eTOF-spectra which is encoded
in two components of $z$, namely $z_{0}$ and $z_{1}$. In the $z_{0,1}$-position
maps it exhibits a dependency resembling sine and cosine functions,
respectively. However, the position is not encoded in a perfect sine-cosine
or circle manner. This is combined to a phase $\phi$ defined by:

\begin{equation}
\phi=\arctan\frac{z_{0}}{z_{1}}\label{eq:Phase}
\end{equation}

\section*{
\begin{figure}[H]
\protect\centering{}\protect\includegraphics[scale=0.45]{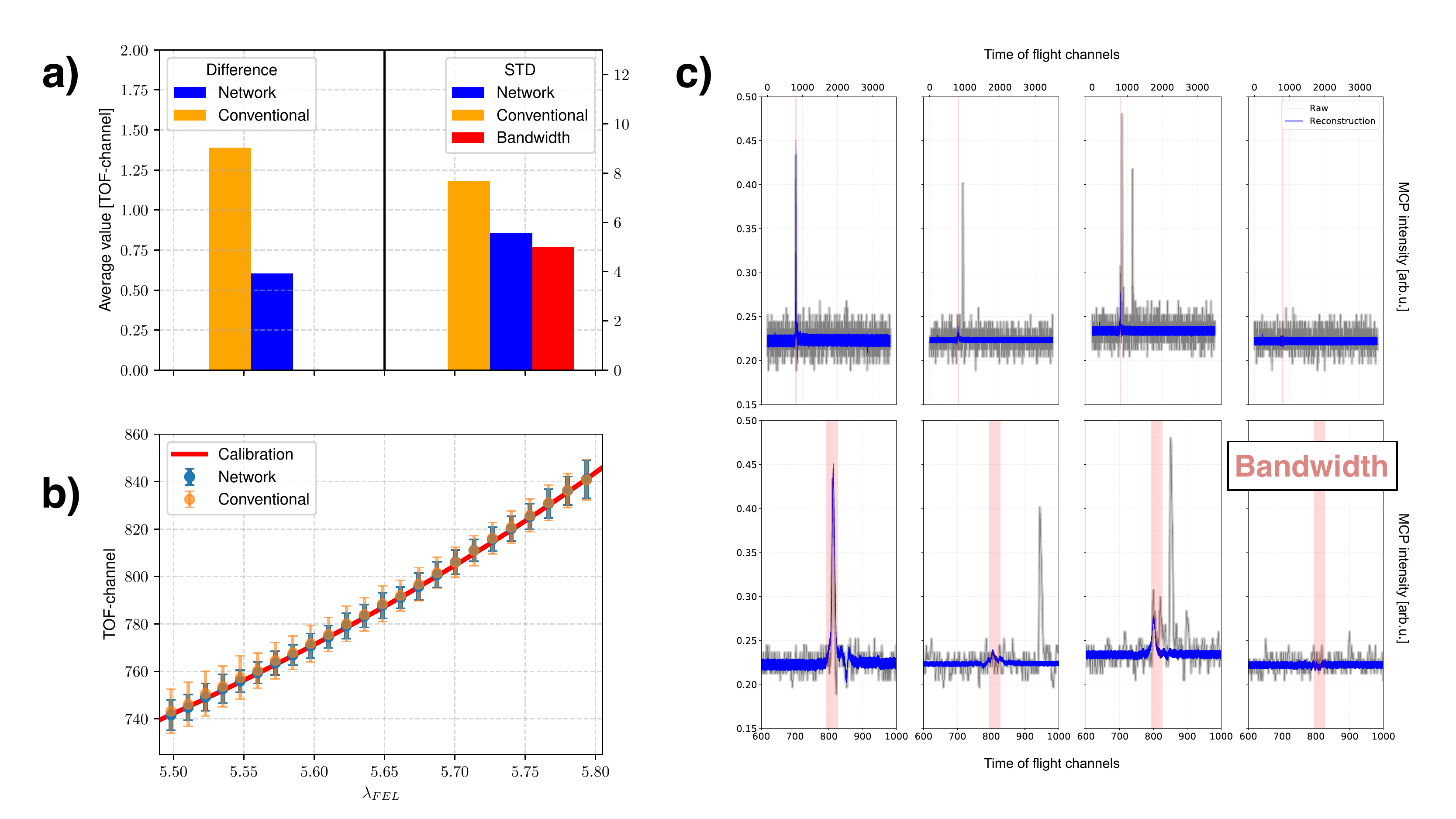}\protect\caption{Comparison of the performance between the traditional analysis and
the neural network: a) The average difference of the network's prediction
of the time-of-flight position (blue) compared with the expected position
at the given $\lambda_{FEL}$ from the calibration is significantly
lower than the average error given by the traditional analysis (orange).
The expected bandwidth is transformed to the STD in TOF channels (red).
The STD from the neural network's predictions are almost identical
to the bandwidth. b) For the 25 $\lambda_{FEL}$, the determined TOF-positions
of the network and of the traditional analysis are compared with a
calibration curve according to the OPIS instrument calibration, which
was independently determined in the instrument commissioning campaigns.
c) An example shot (grey) is shown which has several peaks at different
positions in the eTOFs. The expected position within the given bandwidth
is shown in red. While traditional analysis cannot decide which of
the peaks to designate as real photoelectron signal, the network reconstructs
the peaks at the correct position while ignoring all other peaks in
the raw data.\textcolor{red}{{} }\label{fig:3}}
\protect
\end{figure}
}

In order to provide the most accurate wavelength, $\phi$ is corrected
with an additional neural network. A fraction of the data (3\% high
quality data) where all four eTOFs provide the same information for
the wavelength, i.e. clear photolines at similar positions, is used
to train a fully-connected multi-layer-perceptron (MLP) \cite{MLP}.
This MLP projects $\phi$ to the average least-square fitted TOF-positions
of the 2p peaks from all four eTOFs (see SI). The performance of the
method is evaluated threefold: It is compared to a) the results of
the conventional data analysis (see method section), b) $\lambda_{FEL}$
and c) the center of mass of the magnetic bottle experiment (see SI).
The comparison with $\lambda_{FEL}$ is made by using OPIS' calibration
curve which is shown in Fig.~3b for the network and the traditional
analysis. The results are summed up in Fig.~3a. The average difference
of the network's prediction in TOF-channels is smaller by a factor
of 2 when compared to the conventional method. The estimated bandwidth
of the FEL is translated to a standard deviation value (STD) in TOF-channels.
This STD of the bandwidth is close to the STD of the network's predictions,
whereas the conventional result differs more significantly. To showcase
how the network outperforms the conventional analysis, Fig.~3c depicts
a shot which is difficult to analyse. Multiple peaks with similar
amplitude are appearing at different TOF positions. $\lambda_{FEL}$
including the bandwidth is shown to indicate the region where the
photoelectrons are expected. The network reconstructs the peak in
the correct region, presented in Fig.~3a. Contrastingly the traditional
method struggles to identify the correct peak(s). As an SASE-fluctuation
independent comparison, the predicted wavelength is also compared
to the center of mass of the 2p photoline of sulfur from 2-Thiouracil
in the magnetic bottle experiment, which was running in parallel to
our study. Here, a good agreement is also found and this is presented
in the SI.

Besides the wavelength retrieval, multiple other features are encoded
in the latent space during the unsupervised training process. The
network encodes the intensity distribution of the 4~eTOFs in $z_{2}$,
$z_{3}$ and $z_{4},$ which is plotted in Fig.~2. $B_{1}$ and $B_{2}$
are encoded in two separate components of $z$, namely $z_{4}$ and
$z_{5}$, as shown in Fig.~2. Interestingly, $B_{1}$ only occurs
in two specific bunches of the pulse train and $B_{2}$ is even limited
to only one bunch (see Bunch-No. vs $z_{i}$ maps), indicating synchronised
electronic noise induced from the accelerator environment as a cause.
$B_{1}$ and $B_{2}$ are encoded in an on/off-state and therefore
$z_{4}$ and $z_{5}$ can use an extreme value region for ``on''
and while the baseline disturbance is ``off'' they can use the rest
of the value range for the encoding of a different feature. As a result,
$z_{4}$ also encodes the intensity of eTOF3 while $z_{5}$ is also
encoding $P_{13}$. The network uses the sixth dimension of $z$ for
the other pointing related label $P_{02}$. The linear dependency
of $z_{5}$ vs. $P_{13}$, combined with the cross-like dependency
of $z_{4}$ vs. $P_{13}$, can now be used to determine the variation
of the spatial beam position which can also be an important parameter
for the experiments. $L_{0,1,2,3}$ are fully encoded in $z_{7}$.
The components $z_{8-11}$ are only influencing the reconstruction
in a tiny way and therefore are considered unused. However, reducing
the dimensionality of the latent space increases the overall loss
resulting in a more complicated encoding of the handcrafted labels. 

\section*{
\begin{figure}[H]
\protect\centering{}\protect\includegraphics[scale=0.35]{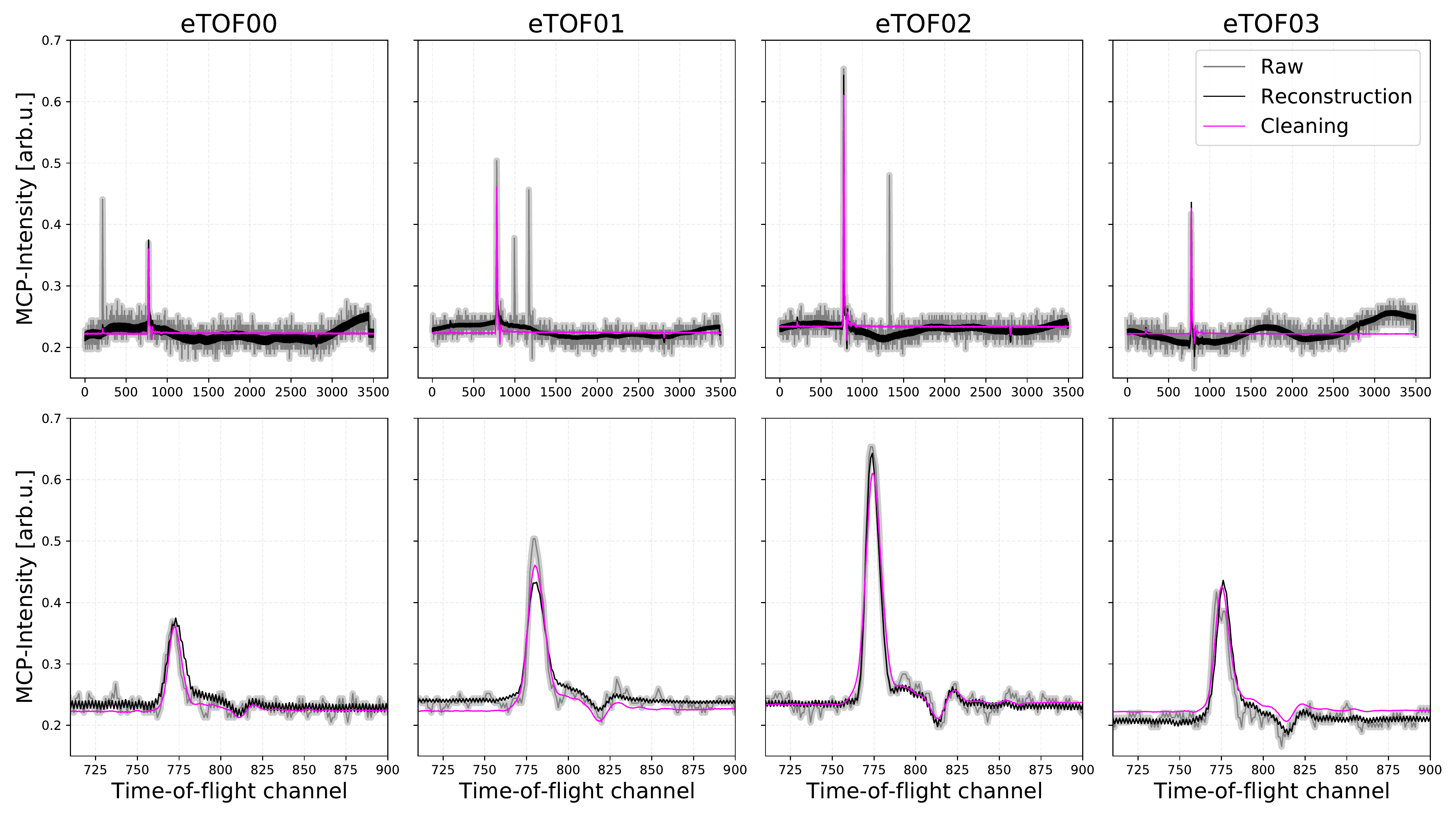}\protect\caption{Data cleaning example: The raw data (grey) is reconstructed via the
network as shown in black. All random hits are discarded, the noise
level is reduced and the prompt signal is reconstructed. A modification
of the latent space allows for eliminating the interleaving issue
and a removal of the baseline disturbance (magenta). \label{fig:4}}
\protect
\end{figure}
}

\subsubsection*{Data cleaning}

The reconstruction of the data by the network alone, already automatically
removes all the random hits from the raw data. Additionally, the noise
level of the baseline is strongly reduced. Finally, with both parts
of the network, the encoder can be used to get the 12D-representation
of the individual samples and consequently one can selectively clean
the compromised data of all these effects, which is shown in Fig.~4.
Since the latent space representation is understood, one can just
change $z_{7}$ from 0.8 (which was determined by the network in order
to achieve the best reconstruction of this specific sample) to the
average value of 0.0 and then by running the decoder with this modified
$z_{7}$ value it is possible to eliminate the interleaving effect.
A similar procedure (see SI) can be used to remove the baseline disturbance. 

\subsubsection*{Outlook}

In order to fully exploit the high-repetition rates of FEL machines
with superconducting accelerators, which deliver FEL radiation with
highly fluctuating photon properties due to the SASE operation mode,
information about the essential parameters are needed on a single-shot
basis. Ideally, this information should be provided by entirely independent
diagnostic devices, which can be operated in parallel to the running
user experiment. This way, the best possible analysis -- even in
a near real-time fashion -- can be enabled, allowing for all possibilities
of data sorting, binning and similar methods, in order to reveal the
dependencies on the photon properties for the physical process under
investigation. This is especially important for photon-hungry experimental
techniques, such as coincidence measurements, which rely on the accumulation
of a large number of single photon interaction events. Blurring or
even disguising dependency effects by averaging data samples covering
a spread of different values of photon properties can be avoided.
OPIS in combination with the trained $\beta$-VAE network can provide
such ability and thus enables the use of the FEL property \textquotedbl wavelength\textquotedbl{}
as an independent sorting parameter for any experimental data analysis.
The next steps will be to train more general networks. The OPIS operation
parameters, i.e. the target gas species, the chamber pressure and
the retarding voltages on the eTOFs, have been kept at fixed values
for the results that are presented in this work. We have recorded
and will record spectra for a variety of combinations of these parameters.
First, dedicated networks will be trained for different operation
parameters. In this case, for each operation mode a specific network
can be used for online analysis. Second, only a single network will
be trained for all operation parameters, allowing the use of the same
network for all operation modes. These two approaches will then be
compared.

\section*{Conclusion}

We have shown that an optimised $\beta$-VAE-network is capable of
finding the underlying core principle of high-dimensional photoelectron
time-of-flight spectroscopy data without any a-priori knowledge in
an unsupervised way. All raw data with low signal-to-noise ratio is
denoised and random hits not correlated to the observed photoionisation
processes are discarded. As a consequence, the reconstructed spectra
are of a much higher quality and in certain cases can very clearly
show photoelectron features which are obscured in the raw data and
cannot easily be processed by conventional analysis methods. The representation
in the latent space covers all the main intrinsic physical properties
of the spectrum, providing direct access to essential information
such as the single-shot FEL wavelength. The inference time of the
trained network is fast and therefore it can be deployed as an onine
tool during photon diagnostics measurement, providing crucial information
for FLASH user experiments in real-time. This will enable or improve
the on-the-fly data analysis which helps to enhance the efficiency
of a beamtime. For instance, by monitoring the data quality in terms
of statistics, for the effect under investigation, one can optimise
the recording time and evaluation of the findings. This concomitant
analysis affords the user the ability to adapt the measurements on
the fly throughout the experimental campaign. Furthermore, any offline
post-experiment data analysis will also benefit from the labels provided
by the $\beta$-VAE-network. In this respect, the ability to isolate
or eliminate certain properties of the data by setting the values
of the VAEs representing those properties to zero may be very useful
for a detailed in-depth analysis of the data set. 

\section*{Methods}

\subsection*{OPIS time-of-flight calibration}

For accurate wavelength measurements with OPIS an instrument calibration
is required. In OPIS commissioning campaigns conversion functions
which assign kinetic energy to measured time-of-flight values have
been empirically determined for each retardation voltage setting.
In these calibration measurements either the photon energy or the
electron kinetic energy was precisely known (eq.~2). This has been
achieved by simultaneous measurements, together with an optical grating
spectrometer as a reference, as well as using the intrinsic calibration
capabilities by means of Auger processes. Auger electrons are emitted
with a fixed kinetic energy corresponding to the difference of the
two electron orbitals involved in the Auger transition and hence can
serve as direct kinetic energy markers in the TOF spectrum. Furthermore,
schemes can be used in which the FEL wavelength is tuned until the
TOF position of a photoline of a particular orbital precisely matches
an Auger line position. This also determines the wavelength and therefore
defines the kinetic energy at the TOF position for other photoelectron
lines in the same spectrum. More detailed information about the OPIS
calibration can be found in Refs.~\cite{OPIS} and \cite{OPIS_PCA}. 

\subsection*{Hyperparameter optimisation}

Table~1 shows the hyperparameter space that was explored while $\sim$700
networks were trained. The batch size, the $\beta$ parameter, the
learning rate and the samples per epoch were tested at a fixed value
as well as within a scheduling process. Apart from assessing the overall
loss, which is a combination of the MSE-reconstruction loss and the
KL-divergence disentanglement loss of the latent space, the evaluation
of the network, with regards to the interpretability of the latent
space with the handcrafted labels, was performed via least-square
fitting as shown in Fig.~2. For the reconstruction loss, absolute
error (AE) and binary cross entropy (BCE) were also tested. The components
of $z$ in Fig.~2 (and the text) are reordered for better readability.
In the case of the stochastic gradient descent optimiser (SGD), the
momentum was tested from 0 to 0.9. The 40 million samples are divided
and randomly shuffled into 40~single hdf5-files each containing one
million samples. 33 of these files are used for training, one million
samples as validation data during the training process, and the remaining
six million to test the trained network afterwards. For data loading
purposes, one epoch is defined as an optimisation step within which
the network processes one file, i.e. one million samples. During training,
the network continues training with the same one million samples for
a fixed number of epochs until the data is replaced by another one
million samples from another file and so on. Memorisation of the data,
with regards to a fixed one million sample portion of the data, is
only observed in very deep networks and also only after a couple of
thousands epochs. Due to this effect, the training data in memory
is replaced every 10~epochs, ensuring that overfitting does not occur,
whilst still allowing for fast data transfer to the GPU which is used
to train the network. An additional indication that this way of training
is not compromising the final result is that no abrupt changes are
observable in the loss function if the data set is replaced after
10~epochs. If the number of epochs for the same data is set to~1,
the process can be interpreted as processing the entire training data
of 33~million samples each 33~epochs. The data was min-max-normalised,
i.e. the 8-bit vertical integer range of {[}0,255{]} was transformed
to float values in the interval {[}0,1{]}.

\begin{table}[H]
\begin{centering}
\begin{tabular}{|l|c|c|}
\hline 
Hyperparamter & Tested & Chosen\tabularnewline
\hline 
\hline 
Batch size & 8-32k & 256\tabularnewline
\hline 
Optimiser & Adam, SGD & Adam\tabularnewline
\hline 
Learning rate & $10^{-2}$ - $10^{-8}$ & scheduled: $10^{-5}$ to $10^{-7}$\tabularnewline
\hline 
Activation function & ReLU, sigmoid, tanh, Mish & Mish\tabularnewline
\hline 
Output activation & None, sigmoid, ReLU, Mish & sigmoid\tabularnewline
\hline 
Samples per epoch & 10k - 33M & 1M\tabularnewline
\hline 
Epochs on same samples & 1 - 10000 & 10\tabularnewline
\hline 
Training data size & 1M - 33M & 33M\tabularnewline
\hline 
$\beta$ parameter & 0 - 10 & 0.034\tabularnewline
\hline 
Layers encoder & 1 - 10 & 5\tabularnewline
\hline 
Layers decoder & 1 - 10 & 4\tabularnewline
\hline 
Max. neurons per layer & 20k & 14k\tabularnewline
\hline 
Latent space dimension & 1 - 20 & 12\tabularnewline
\hline 
Normalisation & None, Min-Max, Std & Min-Max\tabularnewline
\hline 
Reconstruction loss & BCE, MSE, AE & MSE\tabularnewline
\hline 
\end{tabular}
\par\end{centering}
\caption{Hyperparameter optimisation: In order to find the best network, the
shown hyperparameters were varied in the given region.}

\end{table}

\subsection*{Phase correction by the MLP}

The MLP for the phase correction of $z_{0}$ and $z_{1}$ has the
following network architecture 

\begin{equation}
1\rightarrow100\rightarrow60\rightarrow36\rightarrow21\rightarrow12\rightarrow1\label{eq:MLP_phase}
\end{equation}

while the input is the phase and the prediction target is given by
the average TOF-position derived by fitting all 4 eTOF spectra. It
was trained over 2000 epochs with 200k samples, a batch size of 100
and a learning rate of $10^{-5}$, while Mish-activation and the Adam-optimiser
were used. The data was not normalised. The prediction quality was
measured in MSE.

\subsection*{Conventional data analysis}

Multiple methods were tested for processing the single-shot raw data
in a robust and efficient manner. The comparison was made regarding
how well the the data agreed with $\lambda_{FEL}$. The best results
were achieved by an iterative procedure which only analyses the region
of interest, TOF-channels {[}600,1000{]}, corresponding to the zoom-in
region in Fig.~3c. First, a threshold of 0.2 (with respect to the
values shown in Fig.~1) is set to determine all possible peak positions
in all four eTOFs (multiple peaks in one eTOF are possible). These
peak positions are integer values of the maximum position(s). Second,
the peak positions of all detectors are compared. If there is more
than one peak in the same window of 20 TOF-channels for multiple detectors,
further processing of these peaks is performed. Otherwise, if the
amplitude of one peak is higher (by an absolute value of 0.15) then
further processing is only performed on this single peak. If not,
then processing continues on all found peaks. All remaining peak positions
are then optimised by calculating the center-of-mass of the peak (with
floating point precision). Additionally, it was also checked to see
whether other analysis methods, e.g. least-square optimisation fit
routines, could be more suitable. It turns out that there is no advantage
when using these other methods, but they create the disadvantage of
a large increase in computing time. The average value of all determined
peak positions is then taken as a final result.

\subsection*{Acknowledgements}

We acknowledge DESY (Hamburg, Germany), a member of the Helmholtz
Association HGF, for the provision of experimental facilities. Parts
of this research were carried out at FLASH2. G.H. thanks Irina Higgins
for fruitful discussions about the $\beta$-VAE application, interpretation
and presentation. We acknowledge the assistance and support of the
Joint Laboratory Artificial Intelligence Methods for Experiment Design
(AIM-ED) between Helmholtz-Zentrum Berlin für Materialien und Energie
and the University of Kassel. Funded in part by the Innovationspool
of the BMBF-Project: Data-X -- Data reduction for photon and neutron
science. Funded in part by the BMBF-Project: 05K20CBA. We acknowledge
financial support from the Swedish Research Council via the Röntgen
Ångström Cluster (RÅC) Program (No. 2019-06093).

\subsection*{Author contributions}

G.H. and M.B. co-wrote the manuscript with input from all authors.
M.B. designed and commissioned the OPIS instrument. M.B., S.D., F.L.,
K.T. and M.G. performed the experiment to record the used data. G.H.
wrote the $\beta$-VAE-code for this study. F.L. performed the data
analysis of the magnetic bottle data. G.H., G.G., P.F., L.V., D.M.,
F.M., S.D., M.B. and J.V. interpreted and optimised the $\beta$-VAE-network. 

\subsection*{Competing interests}

The authors declare no competing financial interests.

\subsection*{Materials \& Correspondence}

Correspondence and requests for materials should be addressed to G.H.
(email: gregor.hartmann@helmholtz-berlin.de). The datasets used and
analysed during the current study as well as the code for the training
process of the neural network are available from the corresponding
author on reasonable request.
\end{document}